\documentclass{ws-ijmpb}
\usepackage{bm,amsmath}
\usepackage[mathcal]{euscript}
\usepackage[hypertex]{hyperref}
\usepackage{graphicx}
\usepackage[numbers,sort&compress]{natbib}
\graphicspath{{figs/}}
\renewcommand{\vec}[1]{\bm{#1}}
\begin{document}
%
%
\catchline{}{}{}{}{}
%

\title{New Possibilities for Obtaining a Steep Nonlinear Current--Voltage
Characteristics in some Semiconductor Structures}

\author{D. I. Sheka}

\address{National Taras Shevchenko University of Kyiv, Kyiv,
Ukraine}

\author{O. V. Tretyak}

\address{National Taras Shevchenko University of Kyiv, Kyiv, Ukraine\\
tov@univ.kiev.ua}

\author{A. M. Korol}

\address{Department of Physics, National University for Food Technologies, Kyiv, Ukraine\\
korolam@nuft.edu.ua}

\author{A. K. Sen}

\address{TCMP Divn, Saha Institute of Nuclear Physics, 1/AF Bidhan Nagar, Kolkata 700064,
India\\
asokk.sen@saha.ac.in}

\author{A. Mookerjee}

\address{S.N. Bose National Centre for Basic Sciences, 3/JD Salt Lake, Kolkata 700098,
India\\
abhijit@boson.bose.res.in}

\maketitle

\begin{history}
\received{\today} %
\end{history}
%
%

\begin{abstract}
Electronic processes in a semiconductor system consisting of some Resonant
Tunnelling Structures, built in the depletion region of a Schottky barrier, are investigated. It is shown that the Schottky barrier can block or unblock the resonant tunnelling current effectively. Tunnelling processes do reveal the coherent character. Sharp nonlinear current-voltage characteristics are observed on both of the forward and the reverse branches.
\end{abstract}



\section{Introduction}

A lot of various resonant tunnelling structures (RTS) are being investigated
widely now. Interesting results have been obtained with the use of RTS's in
combination with other semiconductor structures. In this paper, we consider a
combination of two structures --- a resonant tunnelling structure and a
Schottky barrier (SB). Such a proposal for embedding a RTS inside the decay or depletion region of a barrier may be found in the literature. It may be noted
that it is not too demanding for the fabrication processes, as well. Such a
barrier is often used as a convenient instrument for studying the
characteristics of various electrical systems, including different types of
RTS's. \citet{North88} have studied the effects associated with the electron
reflection at the semiconductor-metal interface of a Schottky collector.
Resonant-tunnelling spectroscopy of quantum dots has been performed in some
works, e.g., Refs.~\cite{Yoh99,Yung99}. Besides, the SB's may be used for
other purposes, as well. For example, the effect of replacing the usual
collector of a standard double barrier resonant tunnelling diode (DBRTD) by a
Schottky layer has been studied \cite{Konishi93,Smith94}. It was shown in
these works that this could improve the frequency characteristics of the
RTS's. In the case considered in this paper, however, the presence of the SB
plays a fundamentally more important role in the electronic processes under
study. In keeping with the current practice in studying device physics, the
quantum mechanical phase of the carrier wave function and the thermal (phonon) broadening of the resonances have not been considered in the calculations
below, on the assumption that their effects are small enough.

We have shown that an SB may work like a blocking barrier for the resonant
tunnelling current in a system consisting of a metal and a semiconductor with a
double-barrier RTS (DBRTS) placed in the space-charge (or, the depletion)
region. As a result, in some cases one may observe a very sharp (jump-like)
increase, and in some other cases, a similarly sharp jump-like decrease in the
total current through the structure. Thus, we would like to point out that,
besides the usual drop in current that is inherent in such a structure [3], we
focus our attention here to the additional possibility of realizing a steep
nonlinearity in its current-voltage characteristics (IVC). It should be clear
from what follows that, depending on the parameters of the problem, the
indicated nonlinearity may be manifested in both the forward and the reverse
branches of the IVC. Thus, the SB plays here a role akin to a regulator of the
electronic processes in the structure and, in particular, it influences the
observed shape of the $I-V$ curves.

\begin{figure}
\centerline{\includegraphics[width=3.65in]{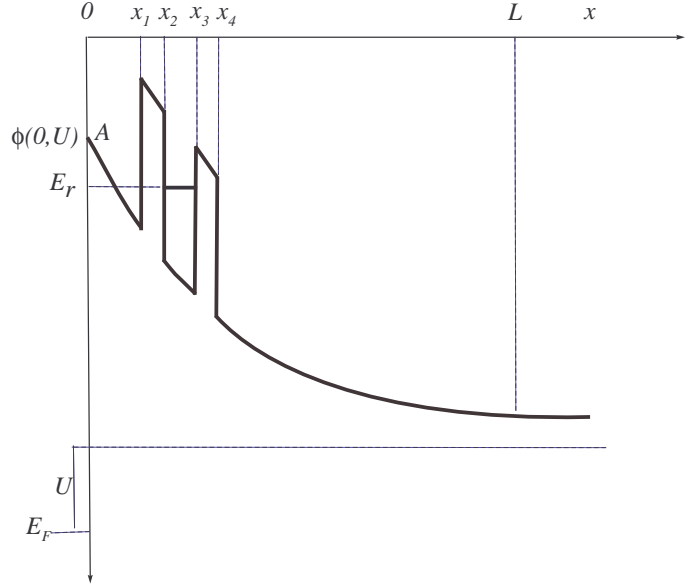}}%
\caption{The potential profile of a RTSC.}\label{fig:1}
\end{figure}

We choose a symmetric double-barrier RTS as the tunnelling structure, and
insert it in the depletion region of a Schottky contact. In the sequel, we
refer to this as a RTSC (Resonant Tunnelling Schottky Contact). The choice of
the double-barrier structure is not fundamental. What is important is that
only the following conditions be met: (i) the RTS chosen must have at least
one resonance level ($E_{r}$) to serve as a channel for the current flow, and
(ii) the size of the RTS must be considerably smaller than the width of the
Schottky layer. Then the magnitude of the $I-V$ response of the RTS must
depend substantially on its starting parameters, e.g., the supposed position
of the resonance level $E_{r}$ in relation to the top of the SB at the zero
voltage bias $U$; $\phi(x=0,U=0)\equiv \phi_0$ (see Fig.~\ref{fig:1}). If
$E_{r}-\phi_0>0$ and $U=0$, then the difference between the indicated levels
increases monotonically, and the IVC has the usual exponentially increasing
form. A qualitatively different situation arises when $E_{r}-\phi_0<0$ and
$U=0$. In that case, the resonance channel for the current flow (that of the
RTS) is initially blocked by the SB; and the current is governed by the
non-resonant tunnelling of electrons and the flow component above the barrier.
As $U$ is increased, the difference $\lvert E_{r}(U)-\phi(x=0,U)\rvert$
decreases, and at a certain voltage $U=U_{c}$, a turning of the resonant
channel of current passage occurs. This turning-on of the channel is
accompanied by an exceedingly sharp rise of the current, and that is reflected
in the shape of the IVC. Thus, there is a substantial difference in the
functioning of a standard DBRTS and the system considered here. In the
standard system, which is initially (i.e., for $U=0$) a symmetric structure,
it is fundamentally impossible to have the conditions for a resonant tunnelling
current. But, the RTSC is an asymmetric structure under an external voltage
(e.g., $U>0$) and a tunnelling current arises naturally.

We point out some additional differences reflected in the IVC between a
standard DBRTS and the structure investigated here. First, the collector and
the emitter regions function somewhat differently. For example, in a standard
DBRTS, the collector plays the role of a reservoir capable of accepting
electrons with any energy (which come from the emitter). In a RTSC (under a
forward bias, $U>0$), the emitter is a bulk semiconductor region, while the
collector is the contact electrode, which can receive only an
energy-restricted fraction of the electrons due to the existence of the
Schottky barrier. Next, one notes that the main electron emission in a
standard DBRTS takes place in the energy interval $[0, E_{F}]$. But, in the
RTSC structure, electrons driven at high energies (due to the external field)
reach the collector. Thus, the \emph{working} energy interval corresponds, in
a certain comparative sense, to \emph{hot} electrons. Ultimately, these
differences give rise to some interesting features in the IVC of the resonant
tunnelling structures.

\begin{figure}
\centerline{\includegraphics[width=3.65in]{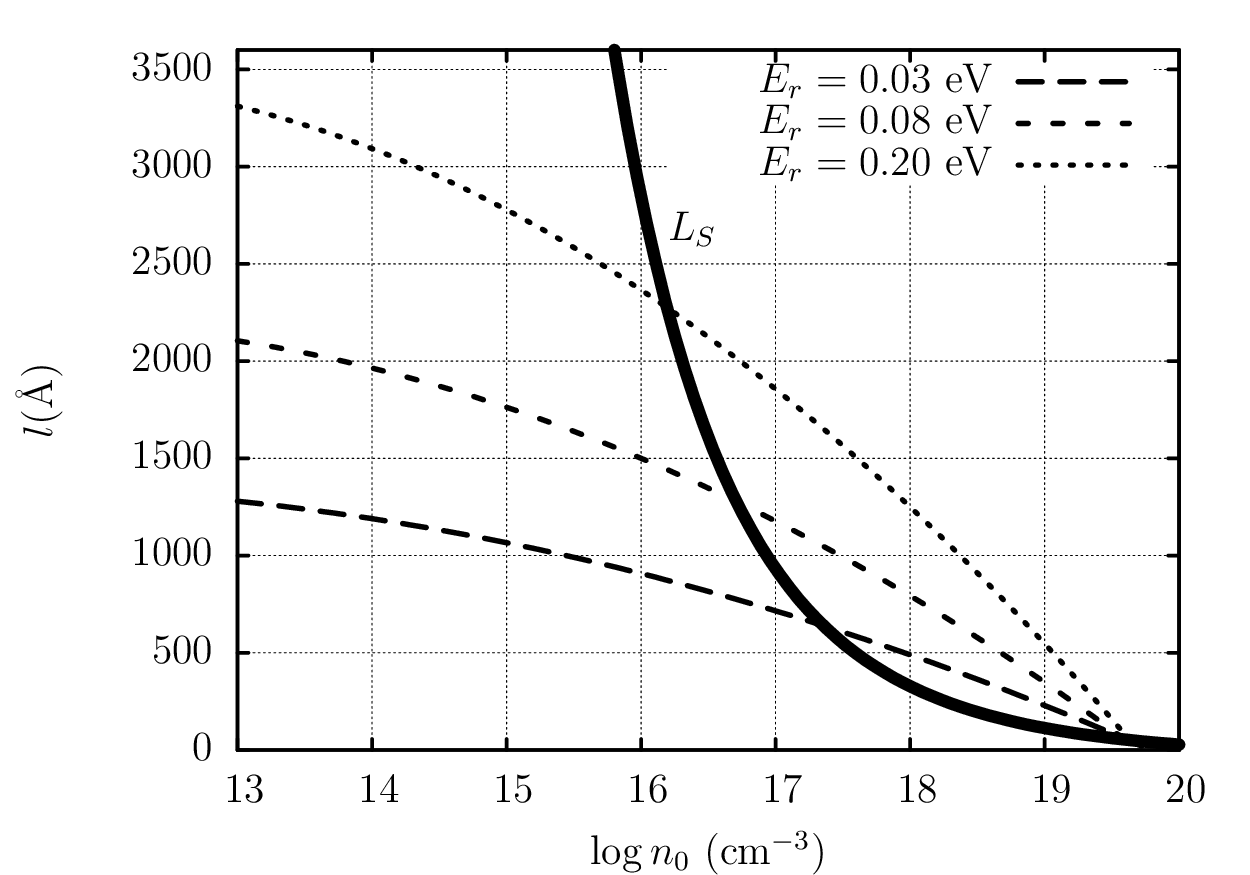}}%
\caption{Dependence of the mean free path ($l$), as well as the width of the
depletion region ($L_S$), on the concentration ($n_{0}$) of shallow doping
impurities.}\label{fig:2}
\end{figure}

The main goal of this paper is to demonstrate the possibility of obtaining
some extraordinarily steep IVC in a modern semiconductor structure. So, for
simplicity, we have omitted from our considerations, such factors as the
accumulation of charge (Coulomb effects), roughness of the surface,
non-parabolicity of the dispersion relation, etc., which should not influence
the effect under study in a qualitative way. We should also point out that the
results obtained in this study are valid in cases when the resonant tunnelling
current exceeds the forward tunnelling current and the current at energies
above the barrier. Our comparison of these currents shows that the stated
condition is well satisfied for a wide spectrum of parameters of the problem.
We calculate the current density with the formula, normally used for
structures such as the RTSC's, namely,
\begin{equation} \label{eq:j} %
j=\frac{emk_B T}{2\pi^2\hslash^3}\int\limits_0^\infty \mathrm{d}E {D\left( {E,U} \right)}
\ln \frac{1+\exp\left[ {\left( {E_F -E} \right)/k_B T} \right]}{1+\exp\left[
{\left( {E_F -E-eU} \right)/k_B T} \right]},
\end{equation}
where $k_{B}$ is the Boltzmann constant, $\hslash=h/2\pi$, $h$ is the
Planck constant, $e$ is the charge of the carrier (an electron here), $m$ is
its effective mass, $E$ is its energy, $T$ is the temperature, $E_{F}$ is the
Fermi level, $U$ is the external potential (bias), and $D(E)$ is the
energy-dependent transmission rate. The quantity $D(E)$ is expressed via the
rates of transparency for the depletion region, $D_{s}(E)$, and that for the
DBRTS, $D_{r}(E)$. If the tunnellings through the DBRTS and the SB are
incoherent (i.e., statistically uncorrelated), $D(E)$ is equal to the product
$D_{s}D_{r}$. This approximation is correct when the mean free path of an
electron ($l)$ in the semiconductor is less than the width of the depletion
region. We note here the fact that the electrons with resonant energies
$E_{r}<0.1$eV only, take part in tunnelling in the structure considered. These
are energies of the order of the lowest resonant level in the DBRTS for
typical parameters. The dependence of the mean free path ($l$) on the
concentration ($n_0$) of doped shallow impurities, at $T=300K$, are shown in
the Fig.~\ref{fig:2}. This dependence is calculated using the data on the
electron mobility for various $E_r$ \cite{Sze81}. The dependence of the width
of the depletion region $L_{S}$ on $n_0$, $L_S =\sqrt {\dfrac{\varepsilon_S
\phi(0)}{2\pi e^2n_0}},$ is also shown in this figure using typical values of
the dielectric permittivity, $\varepsilon_{S}=10.4$ (GaAs), and the height of
the potential barrier at the interface $\phi(0) = 0.6$eV. We note from the
Fig.~\ref{fig:2} that the transmission of the electrons have to be coherent in
the concentration domain of $10^{17}\text{cm}^{-3} < n_0 <
10^{19}\text{cm}^{-3}$.

\begin{figure}
\centerline{\includegraphics[width=3.65in]{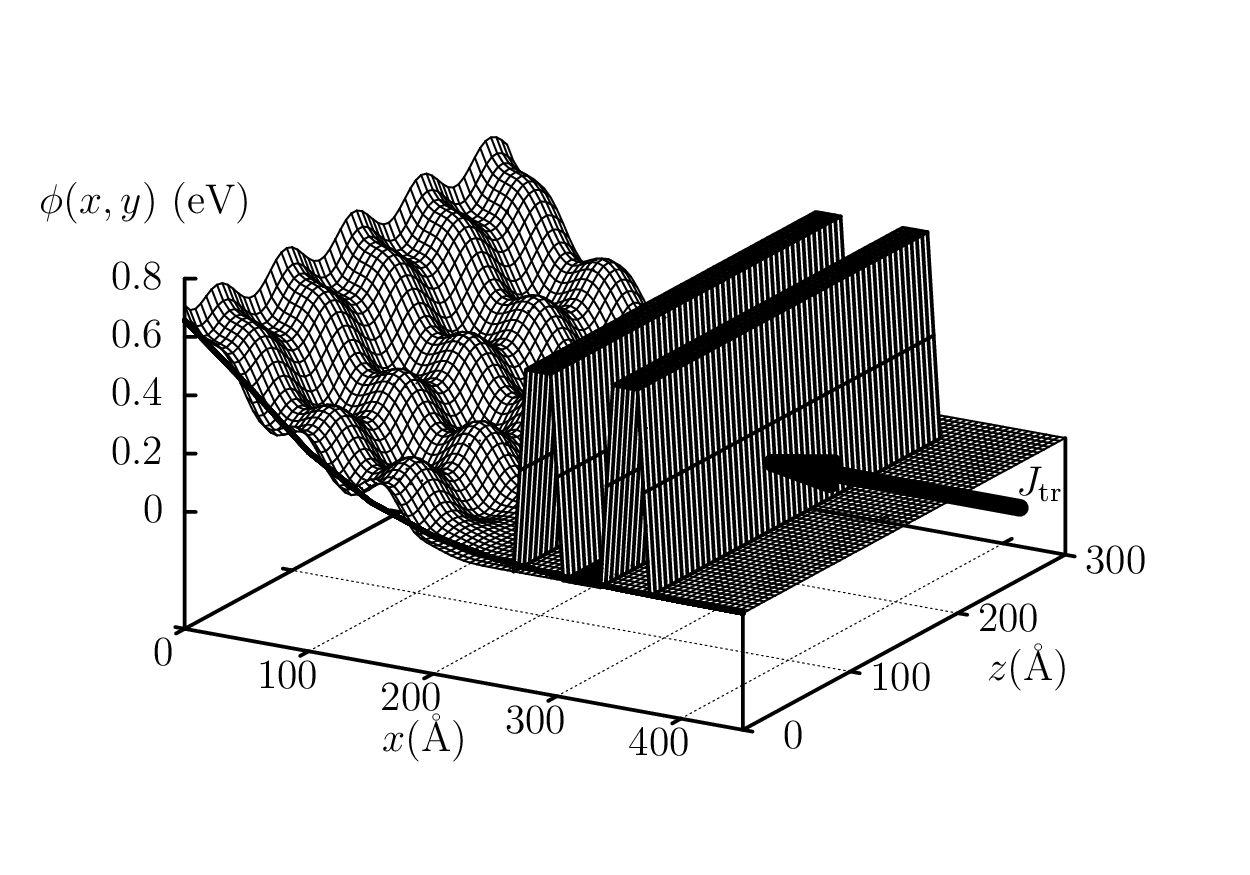}}%
\caption{The shape of a typical potential in the RTSC as a function of the
coordinates $x$ and $z$.}\label{fig:3}
\end{figure}

According to this, we must, in fact, consider the process of the resonant
tunnelling in the three-barrier (two-well) structure in our case. Quantitative
characteristics of this process are strongly dependent on the details of the
potential structure given. We should emphasize here that the Schottky
approximation, widely used for the description of the potential profile
(homogeneously smeared positive charges with concentration $n_0$), is improper
for $n_0\geq 10^{18}\text{cm}^{-3}$, because the width of the Schottky layer
becomes comparable with the distances between the impurity centres,
$a=n_0^{-1/3}$. For this reason, we have justified the model description of a
potential in a depletion region, based on the solution of the Poisson
equation,
\begin{equation} \label{eq:Poisson} %
\nabla^2 \phi \left( {\vec {r}} \right)=-\frac{4\pi e^2}{\varepsilon_S }n\left(
{\vec {r}} \right)\theta \left(\phi \right),
\end{equation}
with the boundary conditions:
\begin{subequations} \label{eq:BC}
\begin{align} \label{eq:BC1} %
\left. {\phi (\vec r)} \right|_{x = 0}  &\equiv \phi (0) = \phi _0  -
eU, \\ %
\label{eq:BC2} %
\left. {\phi (\vec r)} \right|_L  &= \left.\vec{\nabla}_n \phi (\vec r)
\right|_L  = 0,
\end{align}
\end{subequations}
where $\phi_0$ is the height of the potential in the depletion region at zero
bias ($U=0$) and $L(y,z)$ is the surface defined by the Eq.~\eqref{eq:BC2}.
The space distribution of the positive charges has been modelled as follows:
the ``smeared'' unit charges are placed on the sites of a cubic lattice with
the period $a=n_0^{-1/3}$ and the volume $\Omega=(aN)^3$, $N$ being the number
of lattice points in each direction ($x,y,z$). Mathematically stated,
\begin{equation} \label{eq:n}
n(\vec r) = \sum\limits_{\vec l} (\gamma /\pi )^{3/2} \exp \left( - \gamma
(\vec r - a\vec l)^2 \right), \quad l_x ,l_y ,l_z  = \overline{0\dots N}.
\end{equation}
The approximate solution of the Eq.~\eqref{eq:Poisson}, with the boundary
conditions \eqref{eq:BC}, was obtained by using the procedure of minimization
of the variational functional, shown below, with a trial function satisfying
the boundary conditions in Eq.~\eqref{eq:BC}:
\begin{equation} \label{eq:phi}
\phi (\vec r) = \phi(0)\left[1 - \frac{x} {{L(y,z)}}\right]^2\!\!, \quad L(y,z) = \sqrt {\frac{{\varepsilon_s \phi (0)}} {{2\pi e^2 }}} \left[ n_0  + \nu
\sum\limits_{\vec \lambda } e^{ - \mu (\vec \rho  - a\vec \lambda )^2} \right]^{ - 1/2}\!\!\!.
\end{equation}
Here $\vec{\rho} =\left( {y,z} \right)$, $\vec{\lambda} =\left( {l_y ,l_z }
\right)$, and $\mu$, $\nu $ are the variational parameters. The calculated
spatial potential profile, with the built-in DBRTS in the depletion region,
\begin{equation} \label{eq:Phi}
{\Phi (x,0,z)} \Bigr|_{x \geq 0}  =  {\phi (\vec r)} \Bigr|_{z
= 0}  + V_0 \Bigl[ {\theta (x - x_1 )\theta (x_2  - x) + \theta (x - x_3
)\theta (x_4  - x)} \Bigr]
\end{equation}
is shown in the Fig.~\ref{fig:3}, for the following values of the parameters:
$\phi(0)=0.6$ eV, $n_0=4\times 10^{18}\text{cm}^{-3}$, $\gamma = 6.3 \times
10^{12}\text{cm}^{-2}$  (these values define the reduction of the ``point''
charge magnitude by ten times at a distance equal to the nearest centre), and
$x_1=190$\AA (the distance of the nearest barrier of the DBRTS from the
metal-semiconductor interface. The DBRTS parameters are $x_2-x_1 =
x_{4}-x_3=30$\AA (barrier widths), $V_0=1$eV (barrier height), and $x_3-x_2=
40$\AA (quantum well width). The solid line in the Fig.~\ref{fig:3}, refers to
the case $n( {\vec {r}} )=n_0$.

\begin{figure}
\centerline{\includegraphics[width=3.65in]{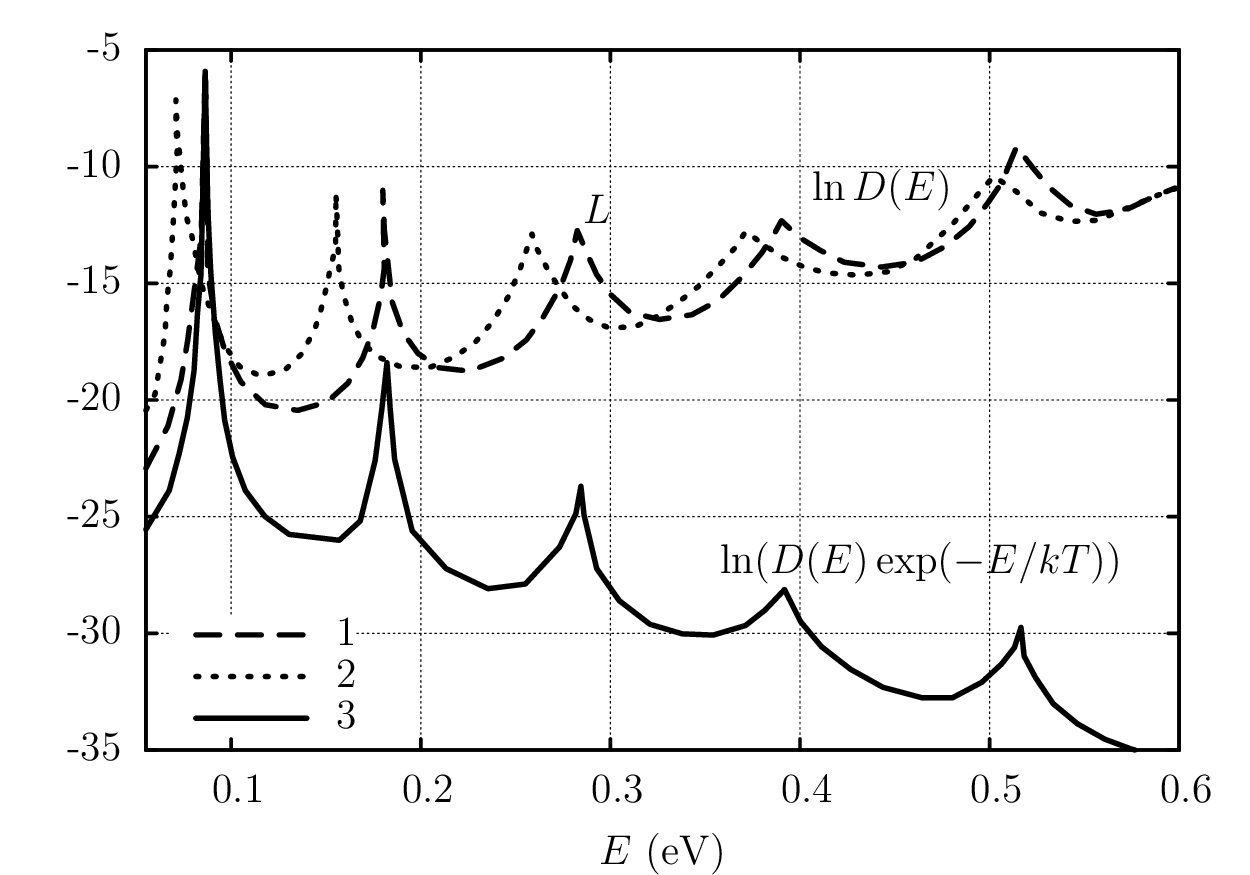}}%
\caption{The function $D(E)$ for two different values of $n_0$: the curve 1
refers to the value $n_0 = 4\times 10^{18}\text{cm}^{-3}$ and the curve 2 to
$n_0 = 6\times 10^{18}\text{cm}^{-3}$. The curve 3 describes the integrand in
Eq.~\eqref{eq:j} at the room temperature ($T=300$K).} \label{fig:4}
\end{figure}

The calculation of the transmission rates $D(E)$ was carried out using the
modified WKB procedure (e.g., see Ref.~\cite{Dobrovolsky98}). In accordance
with the WKB approximation, the wave function of an electron in the region
$x_j  \leq x \leq x_{j + 1}$ is of the form $\psi_j= A_j F_j$, where $A_j  =
\begin{pmatrix}{a_j }\\{b_j }  \\ \end{pmatrix}$,
$F_j  = \begin{pmatrix} \phi_j^- \\\phi_j^+ \\ \end{pmatrix}$, $\phi_j^\pm (x)
= {\left| {k_j (x)} \right|}^{-1/2}\exp \left[\pm \int\limits_{\xi_{j - 1}}^x
k_j (x)\mathrm{d}x \right]$, $a_{j}$, $b_{j}$ are constants, $k_j \left( x
\right)=\sqrt {2 m_j \left( {\Phi (x)-E} \right)}$, $m_{j}$ is the effective
mass in the region $j$, and $\Phi(x)$ is the potential profile of the
structure considered. We have $\xi_{j}=x_{j }$ for the case of the barriers
with vertical walls. The matrices $A_{j}$, which refer to the neighbouring
regions, are connected to each other by $A_{j+1}= G_{j}M_{ j}N_{j}$, where
\begin{equation} \label{eq:mattrices}
\begin{split}
G_j  &=
\begin{pmatrix}
{\frac{1} {2}g_j (1 + g_j^2 )e^{i(k_j  - k_{j + 1} )x_j } } & {\frac{1}
{2}g_j (1 - g_j^2 )e^{ - i(k_j  + k_{j + 1} )x_j } }  \\[2pt]
{\frac{1} {2}g_j (1 - g_j^2 )e^{i(k_j  + k_{j + 1} )x_j } } & {\frac{1}
{2}g_j (1 + g_j^2 )e^{ - i(k_j  - k_{j + 1} )x_j } }  \\
\end{pmatrix},\\
g_j  &= \left( {\frac{{m_{j + 1} }} {{m_j }}} \right)^{1/4}, \quad M_j  =
\begin{pmatrix}
e^{ - \delta _j } & 0  \\
0 & e^{\delta _j }  \\
\end{pmatrix},\\
N_j  &=
\begin{cases}
T, & \text{when }{\left.\dfrac{\mathrm{d}U}{\mathrm{d}x}
\right\vert_{\xi _{j + 1} }  > 0}  \\
{T^+}, & \text{when }{\left.\dfrac{\mathrm{d}U}{\mathrm{d}x}
\right\vert_{\xi _{j + 1} }  < 0}  \\
\end{cases}
,\quad %
T =
\begin{pmatrix}
{\frac{1} {2}e^{i\pi /4} } & {\frac{1}{2}e^{ - i\pi /4} }  \\
{e^{ - i\pi /4} } & {e^{i\pi /4} }  \\
\end{pmatrix}.
\end{split}
\end{equation}
Here $\delta _j =\int\limits_{\xi _j }^{\xi _{j+1} } {k_j (x)} \mathrm{d}x$,
and we use the Jeffrey's transformations (e.g., see Ref.~\cite{Froman65}) to
write the matrix $T$. The matrices $G_{j}$ were obtained from the condition of
continuity of both the wave functions and the flux must be continuous at
$x=x_{j }$. The rate of transparency is defined as
\begin{equation} \label{eq:D}
D = \left[\left(\prod_{j = 1}^S G_j M_j N_j \right)_{11} \right]^{-2}
\end{equation}
for a structure which incorporates a number of $S$ interfaces. The result for
the calculation of $D(E)$ is shown in the Fig.~\ref{fig:4}, for the same
parameters as for Fig.~\ref{fig:3}, alongwith $\gamma=0$ and $U=0.1$eV. It is
seen clearly that the main contribution to $D(E)$ is due to the first (lowest
in energy) resonant level $E_{r}$ which lies energetically close to the
resonant level of the DBRTS.

\begin{figure}
\centerline{\includegraphics[width=3.65in]{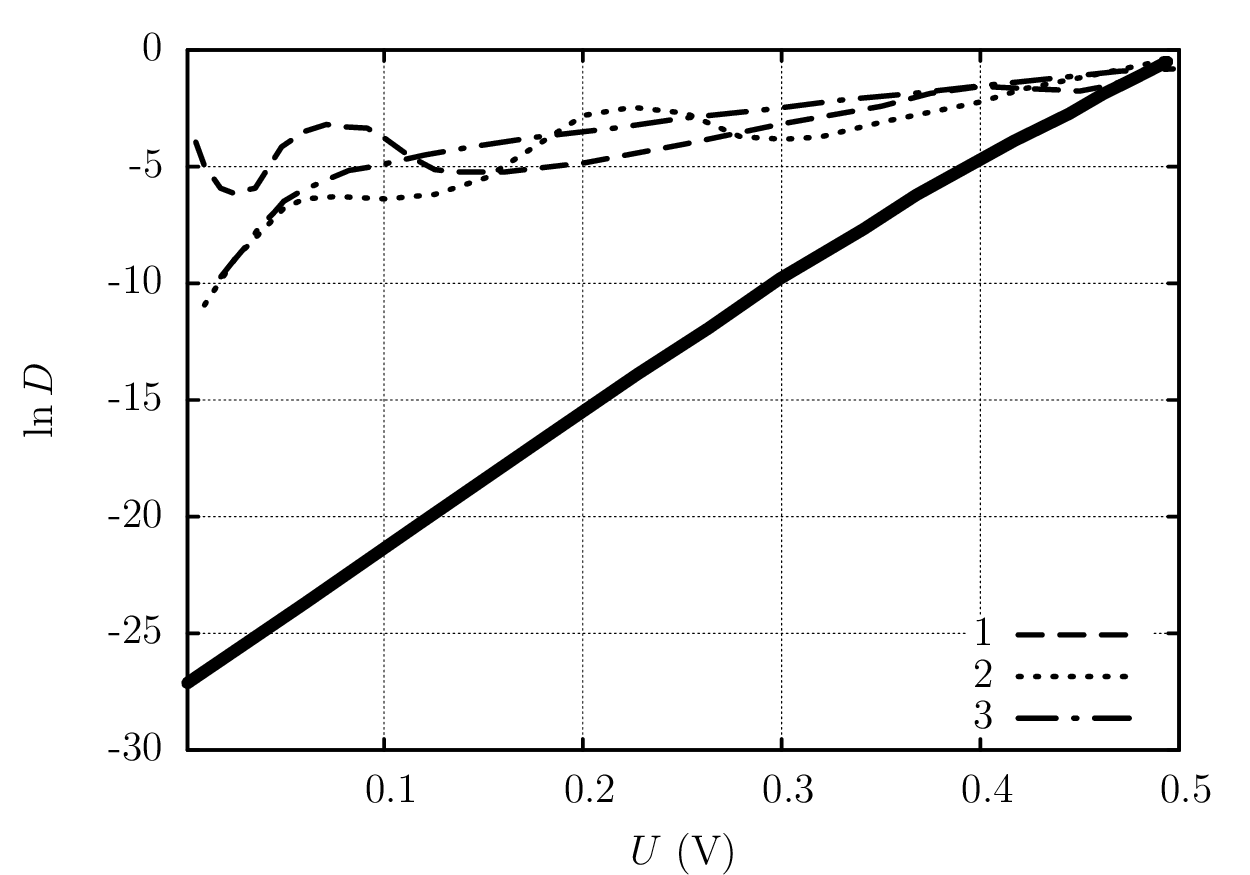}}%
\caption{Dependence of $\ln D(E_{r})$ on $U$ for the minimum (curve 1) and the
maximum (curve 2) values of $L$. The curve 3 represents the surface
[$L(y,z)$]-averaged dependence of $\ln D(E_{r})$ on $U$. The solid line refers
to the case of incoherent tunnelling of electrons through all the barriers of
the RTSC.} \label{fig:5}
\end{figure}

The coordinate dependence of the depletion region $L(y,z)$,
Eq.~\eqref{eq:phi}, results into a significant dependence of the transmission
rates on the coordinates of the \emph{percolation point} $(y,z)$ on the
surface described by Eq.~\eqref{eq:phi}. In the Fig.~\ref{fig:5}, we show the
dependence of $D(E_{r})$ on $U$ for the minimum (at $y=z=a/2$, red curve) and
the maximum (at $y=z=$0, blue curve) respectively of $L(y,z)$. The green curve
in this figure represents the surface-averaged [over $L(y,z)$] dependence of
$\left.D_{\text{ef}}(E)\right\vert_{E=E_r}$ on $U$. The black line refers to
the case of incoherent tunnelling through all the barriers of the RTSC.

To evaluate the current we have to determine the quantity $\Phi(x)$, which
consists of the two terms: the potential of the depletion region $\phi(x,U)$
and the potential energy of the DBRTS. The first one is defined as the
solution of the Poisson equation, see Eq.~\eqref{eq:Poisson},
\begin{equation}\label{eq:Poisson2}
\Delta \phi  =  - \frac{{4\pi e}} {{\varepsilon _s }}\rho (x),\quad \rho =
  \begin{cases}
    en_0 , & 0 < x < x_1 ,\quad  x_4  < x < L, \\
    0 & x_1  < x < x_4 ,\quad x > L.
  \end{cases}
\end{equation}
with the usual boundary conditions for the depletion region of width $L$ [cf.
Eq.(3)]
\begin{equation} \label{eq:BC3}
\phi (0,U) = \phi(0,0) + eU, \quad \phi (L,U) = \left. {\frac{{\partial \phi
(x,U)}}{{\partial x}}} \right|_{x = L}  = 0.
\end{equation}
As a result, one obtains
\begin{align} \label{eq:phi-result}
\phi (x,U) &= \frac{{2\pi n_0 Ne^2 }} {{\varepsilon _s }}(x - L)^2\nonumber\\
& + \frac{{2\pi n_0 Ne^2 }} {{\varepsilon _s }}
\begin{cases}
{(x_4  - x_1 )(2x - x_4  - x_1 ),} & {0 < x < x_1} \\
{ - (x - x_4 )^2 ,} & {x_1  < x < x_4}  \\
0 & {x_4  < x < L}
\end{cases},\\
\label{eq:L-result} %
L^2  &= \frac{{\phi (0,U)\varepsilon _s }} {{2\pi e^2 }} + (x_4^2  - x_1^2 ).
\end{align}
The values $U>0$ refer to the forward bias (the DBRTS is located in the
coordinate interval $x_1<x<x_4)$. The field intensity in this interval is
defined as:
\begin{equation} \label{eq:F}
F = -\frac{4\pi n_0 e}{\varepsilon _S }\left( {L-x_4} \right).
\end{equation}
The quantity $D_{s}(E)$ necessary for evaluating the current is determined as
a transparency of the barrier with the potential energy in
Eq.~\eqref{eq:phi-result}, plus the potential energy associated with the image
forces $\phi_{\text{im}} =-e^2/(4\varepsilon_s x)$. In the interval of
energies close to the top of the resulting barrier, where the coordinate
dependence has the parabolic form, the transparency coefficient is of the
following form:
\begin{equation} \label{eq:D-result}
D_s (E)\approx \frac{1}{1+\exp\left[\left(\bar{\phi}-E\right)/E_0\right]},
\end{equation}
where $\bar{\phi} = \phi(0,U)-1/(2\beta^2)$, $\beta  =
\left\{\varepsilon_S/[4F(0)] \right\}^{1/4}$, $E_0  = \dfrac{{\hslash
\varepsilon _s }} {{2\pi \sqrt {2m_s } \beta ^3 e^2 }}$, $F(0) =  -
e^{-1}\left. {\dfrac{{\partial \varphi (x,U)}} {{\partial x}}} \right|_{x = 0}
$. Note that $D_{S}(E)$ in Eq.~\eqref{eq:D-result} is a $\theta$-like function
with a half-width $\Gamma =\ln (3+2\sqrt 2 )E_r $, which may be expressed via
the quantities $\delta_{j}$. Evaluation of the current-voltage characteristics
(IVC) using the above formula yields the following expression
\begin{equation} \label{eq:j_r} %
\begin{split}
j_r &\approx \frac{m\Gamma D_1 D_2 }{\pi ^{3/2}(D_1 +D_2 )^2}\left[1 - \exp
\left(-\frac{eV}{k_B T}\right) \right]\exp \left[ {\frac{E_F -E_1 }{k_B T} +
\left(\frac{\Gamma}{4k_B T} \right)^2} \right]\\
&\times \text{erfc}\left\{ \frac{2}{\Gamma }\left[ {\phi (0)-E_1 +\frac{\Gamma
^2}{8k_B T}} \right] \right\},
\end{split}
\end{equation}
where $\Gamma $ is the half-width of the resonant level, and $D_{1}$, $D_{2}$
are the transmission rates for the DBRTS barriers (the explicit forms of the
quantities $\Gamma $, $D_{1}$ and $D_{2 }$ are somewhat cumbersome, and hence
omitted here). This expression is proper for the forward bias as well as for
the reverse one.

\begin{figure}
\centerline{\includegraphics[width=3.65in]{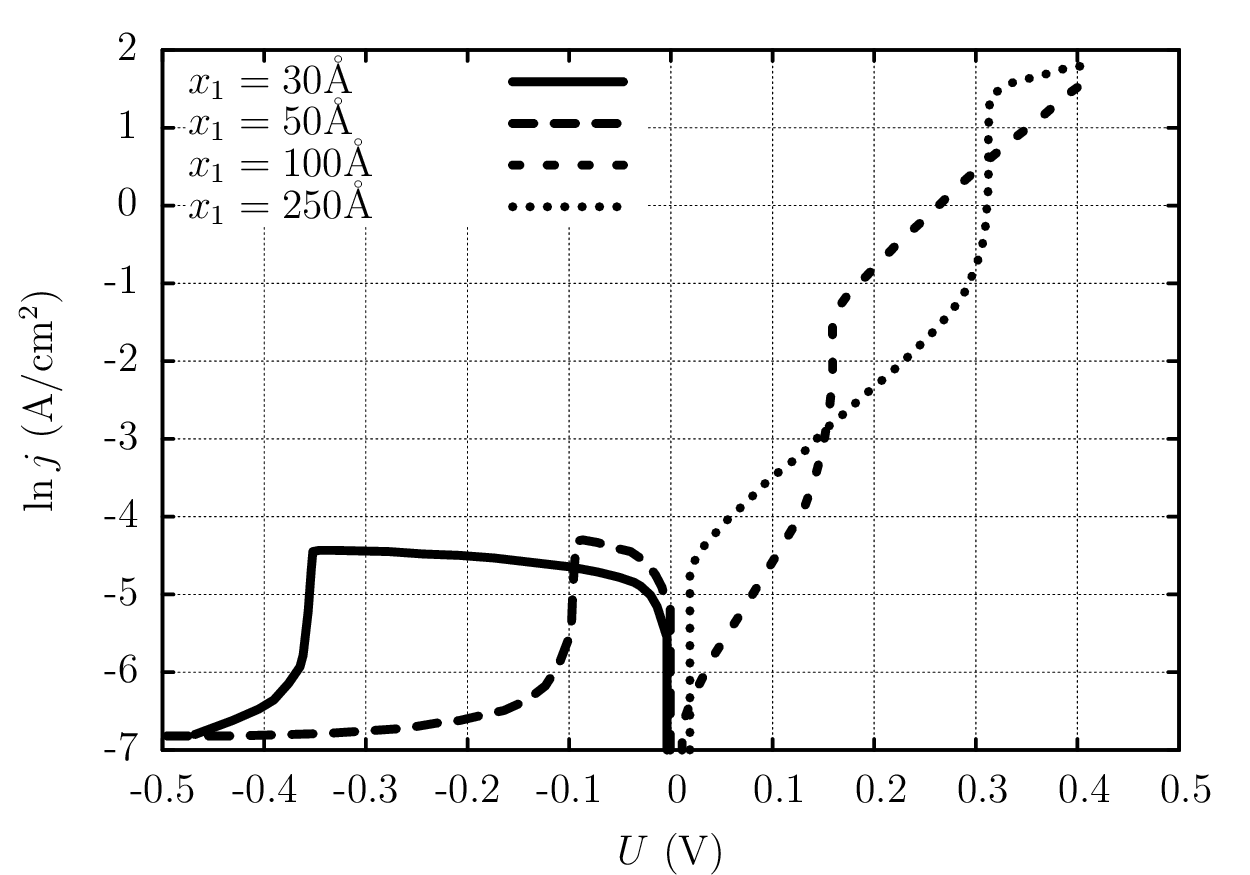}}%
\caption{The current-voltage ($I-V$) characteristics of the RTSC for the
direct and the reverse bias, for four different values of $x_{1}$ (the metal
to DBRTS distance).}\label{fig:6}
\end{figure}

The Fig.~\ref{fig:6} demonstrates both the branches of the IVC for the
following set of parameters at the room temperature ($T=300$K: the DBRTS
parameters the same as mentioned above, $m_{s} = 0.067 m_0$, $m_{b} = 0.1
m_0$, $n_0= 10^{17}\text{cm}^{-3}$ and for four different values of $x_1$ (the
distance between the DBRTS and the metal interface). The IVC's calculated
(using the standard procedure) and shown in the Fig.~\ref{fig:6} include not
only the resonant-tunnelling current but also the over-barrier part of the
current. Some regions with negative differential resistance (or, conductance)
are observed in the back-biased branch of the IVC. The nature of these regions
is obvious: they appear due to the blocking of the resonant-tunnelling current
by the top part of the SB at voltages below a certain voltage $U_{c}<0$. We
would like to emphasize that the peak to valley ratio is of order of $10^2\div
10^3$ in a wide range of parameters involved in the DBRTS.

\begin{figure}
\centerline{\includegraphics[width=3.65in]{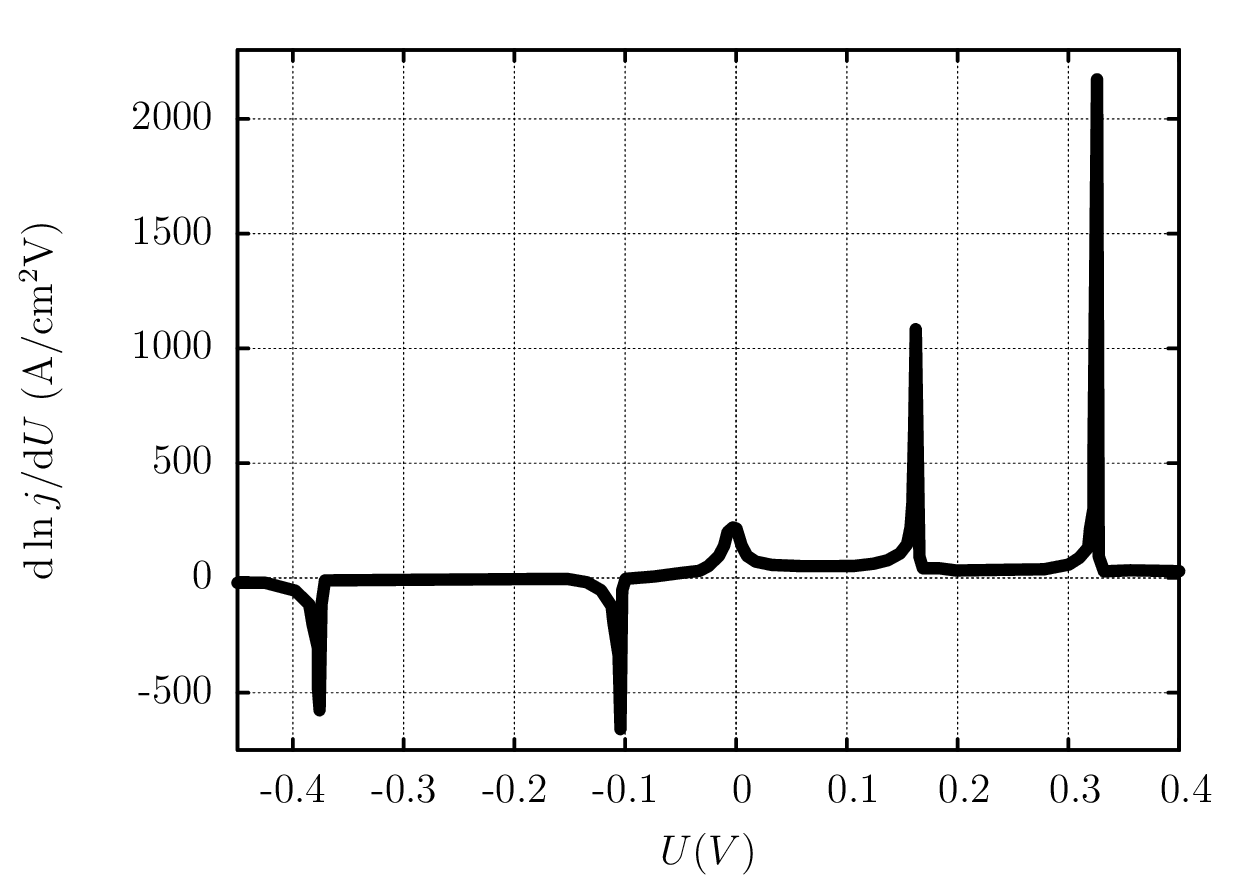}}%
\caption{The dependence of the differential steepness parameter of the IVC,
$\alpha = \mathrm{d}\ln j/\mathrm{d}U$, at $T=300$K, for the same values of
the DBRTS parameters and four values of $x_{1}$ (the metal to DBRTS distance),
the same as in the Fig.~\ref{fig:6}.}\label{fig:7}
\end{figure}

Let us consider the forward branch of the IVC. The plotted functions are
completely consistent with the assumptions made in this paper as to the
character of the $I-V$ curves. At voltages $U$ less than $U_{c}$, the currents
in the investigated RTSC structure are relatively small. Then, in the vicinity
of $U=U_{c }$, there is a precipitous rise in the current. For example, the
current increases by approximately a factor of ten in response to a voltage
change of $0.01$eV. One is also struck by the large values of the parameter
$\alpha =\mathrm{d}\ln j/\mathrm{d}U$, describing the differential steepness
of the IVC. They are much greater than the values of $e/k_{B}T$, typical for
Schottky barriers (see Fig.~\ref{fig:7}). The values of $U_{c}$ and $\alpha $
depend on many parameters of the structure; namely, the height of the SB, the
dopant concentration, the geometric parameters of the DBRTS and its distance
from the metal interface, etc.

To explain the substantial growth of the current as the distance $L$ from the
metal to the DBRTS is increased, it is convenient to refer to the
Fig.~\ref{fig:1}. We see that at large $L$, the opening of the channel for the
resonant tunnelling current (resonant energy $E_r)$ occurs at a higher voltage
$U=U_c$, which corresponds to a lower barrier height $\phi(x=0,U)$.
Consequently, the distribution function makes for higher currents in this
case.

Finally, we should say a few words about the possible advantages of the system
investigated here from the standpoint of practical applications. We note first
that it retains the advantage that motivated the proposal made in the
Refs.~\cite{Konishi93,Smith94} of replacing the conventional collector of the
standard resonant-tunnelling diode by a Schottky collector; thus using the
possibility of reducing the emitter-collector capacitance. Here increasing the
distance between the metal and the DBRTS leads not only to a decrease in
capacitance but also to a simultaneous increase in the steepness of the IVC
(see the Figs.~\ref{fig:6} and \ref{fig:7}).

In addition, it should be noted that possible devices using the IVC given
above (e.g., switches, amplifiers, rectifiers etc.) should possess good
characteristics not only at low temperatures but even around room temperatures
(note that the curves in the Figs.~\ref{fig:6} and \ref{fig:7}, were
calculated for $T=300$K).

\section*{Acknowledgements}

The work presented above, grew out of a bi-lateral Indo-Ukrainian
collaboration during the recent past, under the auspices of the Ministry of
Sciences, Government of Ukraine, and the Department of Science and Technology
(DST), Government of India. The last four authors express their respect and
indebtedness to the (late) first author, Professor Dmitri I. Sheka, for his
insights in this area of physics and dedicate this last paper bearing his
name, to the academic zeal that characterized his persona. We are also
grateful to Denis D. Sheka for helping us with the manuscript.


\end{document}